\documentclass[aps,prb,twocolumn,floatfix,superscriptaddress]{revtex4-1}
\usepackage{graphicx,txfonts,bm}
\usepackage[colorlinks,citecolor=magenta,linkcolor=blue]{hyperref}

\begin{document}


\title{Novel correlated 5$f$ electronic states in cubic $An$Sn$_3$ ($An$=U, Np, Pu) intermetallics}


\author{Haiyan Lu}
\email{hyluphys@163.com}
\affiliation{Science and Technology on Surface Physics and Chemistry Laboratory, P.O. Box 9-35, Jiangyou 621908, China}

\author{Li Huang}
\affiliation{Science and Technology on Surface Physics and Chemistry Laboratory, P.O. Box 9-35, Jiangyou 621908, China}

\date{\today}


\begin{abstract}
The intricate interplay between itinerant-localized 5$f$ states and strongly correlated electronic states have been systematically investigated in isostructural actinide compounds $An$Sn$_3$ ($An$=U, Np, Pu) by using a combination of the density functional theory and the embedded dynamical mean-field approach. The obvious narrow flat 5$f$ electronic band with remarkable spectral weight emerges in the vicinity of Fermi level for three compounds. Subsequently the significant hybridization between 5$f$ states and conduction bands opens evident gaps together with conspicuous valence state fluctuations jointly indicating the partially itinerant 5$f$ electrons. Especially, prominent quasiparticle multiplets only appear in PuSn$_3$ due to the sizable valence state fluctuations and multiple competing atomic eigenstates. Therefore itinerant 5$f$ states tend to involve in active chemical bonding, restraining the formation of local magnetic moment of actinide atoms, which partly elucidates the underlying mechanism of paramagnetic USn$_3$ and PuSn$_3$, as well as itinerant-electron antiferromagnetic NpSn$_3$. 
Correspondingly, the 5$f$ electronic correlation strength expressed in band renormalization and electron effective masse intertwines with itinerant-localized 5$f$ states. Consequently, detail electronic structure of 5$f$ states dependence on actinide series shall gain deep insight into our understanding of $An$Sn$_3$ ($An$=U, Np, Pu) intermetallics and promote ongoing research.
\end{abstract}


\maketitle

\section{Introduction\label{sec:intro}}

The bewildering 5$f$ electron and its hybridization with ligand states ($f-c$ hybridization) interrelate the advent of unprecedented quantum phenomena including complex magnetic ordering, heavy-fermion behavior and unconventional superconductivity~\cite{PhysRevLett.108.017001,PhysRevLett.55.2727,PhysRevLett.37.1511,Sarrao2002,RevModPhys.81.235,shim:2007}, to name a few. Unlike the mostly localized 4$f$ states of lanthanides, the extended 5$f$ electronic wave function produces partially itinerant 5$f$ states. The itinerant-localized dual nature of 5$f$ electrons ubiquitously exists in actinides which enables the tunability of itinerant degree of freedom with enlarging atomic series. Generally, 5$f$ electrons are believed to be mainly itinerant until plutonium (Pu) which locates on the boundary between itinerant and localized 5$f$ electrons~\cite{RevModPhys.81.235}. On account of the narrow energy band of 5$f$ electrons, the 5$f$ states is easily affected by external temperature, pressure, chemical doping and magnetic field. To unveil the long-standing issue of itinerant-localized 5$f$ electrons and their relationship with $f-c$ hybridization, binary $An$Sn$_3$ ($An$=U, Np, Pu) compounds are established as an ideal archetype system to explore the subtle electronic structure and related exotic properties.

\begin{figure}[ht]
\centering
\includegraphics[width=0.9\columnwidth]{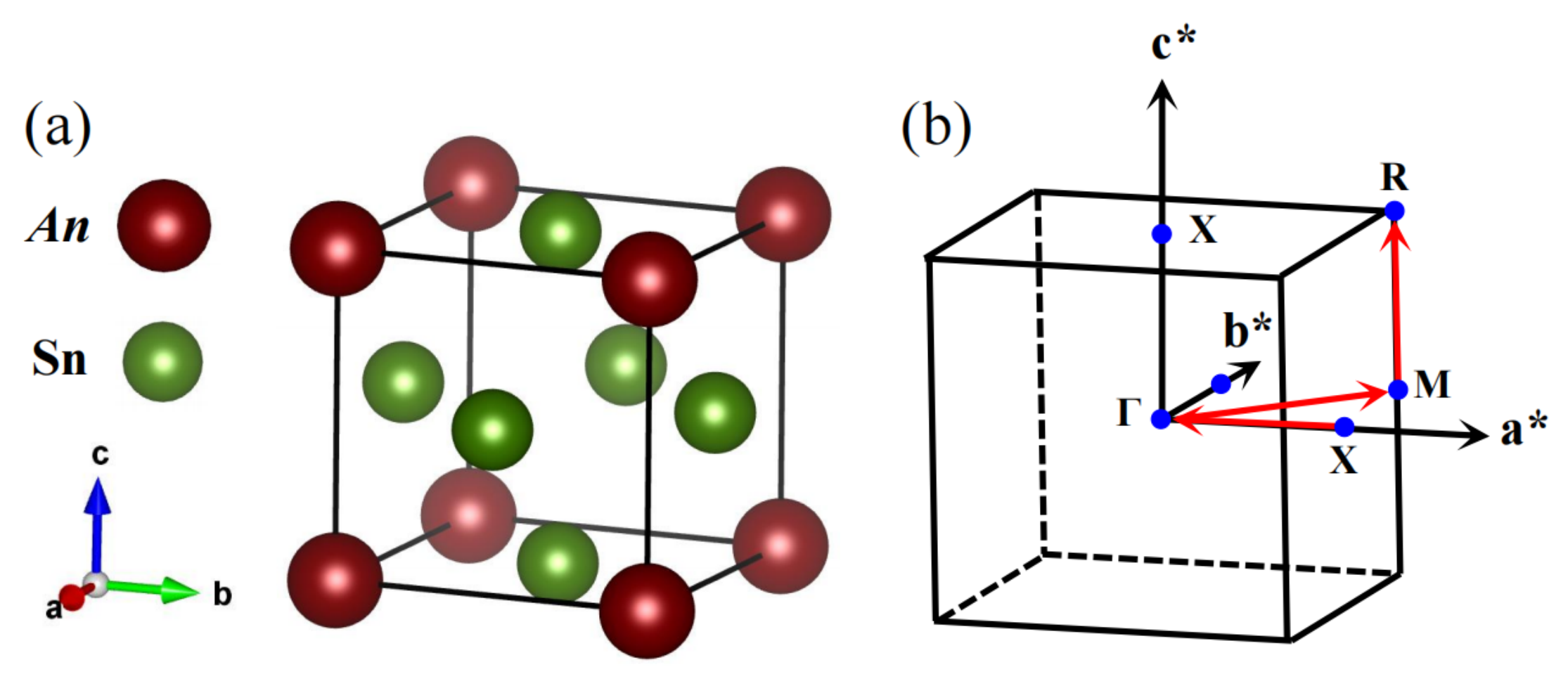}
\caption{(Color online). (a) Crystal structure of $An$Sn$_3$ ($An$=U, Np, Pu). (b) Schematic picture of the first Brillouin zone of $An$Sn$_3$ ($An$=U, Np, Pu). Some high-symmetry $k$ points $X$ [0.5, 0.0, 0.0], $\Gamma$ [0.0, 0.0, 0.0], $M$ [0.5, 0.5, 0.0], and $R$ [0.5, 0.5, 0.5] are marked.
\label{fig:tstruct}}
\end{figure}

$An$Sn$_3$ ($An$=U, Np, Pu) intermetallic compounds crystalize in cubic AuCu$_3$ structure (space group $Pm$-3$m$) [see figure~\ref{fig:tstruct}(a)] with $An$ atom locating at cubic corners and Sn atoms occupying the center surface. As is listed in table~\ref{tab:param}, the experimental lattice constants are 4.603 \r{A}~\cite{PRB31}, 4.627 \r{A}~\cite{cubic237NpM3} and 4.630 \r{A}~\cite{SARI1983301} for USn$_3$, NpSn$_3$ and PuSn$_3$, respectively. Obviously, lattice constants of three isostructural compounds are approximately 4.6 \r{A} and slightly stretch from USn$_3$ to PuSn$_3$.
Since the interatomic distance $d_{An-An}$ is much larger than the Hill limit ($\sim$3.5 \r{A})~\cite{Hilllimit}, these intermetallics with weak 5$f$ electrons overlap are expected to develop certain magnetic ordering at low temperature. However, some exceptions~\cite{PhysRevB.41.9294,PhysRevB.39.13115} with paramagnetic ground state implicate the crucial role of hybridization between 5$f$ electron and ligand states, which is also tightly correlated with the heavy-fermion behavior~\cite{PhysRevB.33.8035,PhysRevLett.37.1511,LuPuSn3} of $An$Sn$_3$ ($An$=U, Np, Pu).
.

USn$_3$ is a heavy-fermion compound with an electronic specific heat coefficient 171 mJ/(mol$\times$K$^2$)~\cite{PhysRevB.33.8035}, which has been verified by the temperature dependence of electronic specific heat coefficient~\cite{PhysRevB.33.8035} and electronic resistivity experiments~\cite{PhysRevB.9.1381}.
Peculiarly, USn$_3$ remains paramagnetism at very low temperature 1 K with a single weak de Haas–van Alphen (dHvA) frequency at 0.5 K~\cite{PhysRevB.59.14473}. The paramagnetic ground state has been elucidated by the 5$f$ orbitals hybridization with ligand states which induces strong spin fluctuations observed by inelastic neutron scattering study~\cite{PhysRevB.41.9294}. 
Since USn$_3$ approaches the verge of magnetic instability, a small doping with Pb may initiate antiferromagnetic ordering~\cite{PhysRevB.41.9294}. The physical picture is consistent with a spin density wave magnetic unstability at the quantum critical point detected by using $^{119}$Sn nuclear spin-echo decay rate measurement~\cite{PhysRevLett.102.037208}. 
Since the heavy-fermion behavior and paramagnetism are closely associated with the subtle electronic structure, x-ray photoemission spectroscopy of USn$_3$ detects a significant peak around the Fermi level which is contributed by 5$f$ states, as well as a small shoulder consisted of 6$d$-5$p$ hybridization, accompanied by a hump at about 7 eV from Sn-5$s$ states~\cite{PhysRevB.35.7922}. Nevertheless, the extremal orbits predicted by previous first-principles caculation~\cite{Kadowaki507} have not been experimentally corroborated. The reason my arise from the heavy electrons which are substantially renormalized by many-body interactions could not be correctly described with single-particle band theory. 
Moreover, angle-resolved photoemission spectroscopy experiment which unravel the fine 5$f$ electronic states are still deficient.

NpSn$_3$ is the first itinerant-electron antiferromagnetic compound with colinear antiferromagnetic ordering and a small magnetic moment 0.28 $\mu_B$~\cite{PhysRevLett.37.1511,PhysRev.143.245}. Similar to USn$_3$, the atomic distance between two nearest Np atoms exceeds Hill criterion~\cite{Hilllimit} indicating the basically localized 5$f$ electrons.
Abundant experiments have envinced the localized behavior of 5$f$ electrons comprising the deduction of electronic specific heat coefficient from paramagnetic state 242 mJ/(mol$\times$K$^2$) to 88 mJ/(mol$\times$K$^2$) below N\'{e}el temperature 9.5 K~\cite{PhysRevB.33.3803}, temperature dependent electronic properties~\cite{PhysRevLett.65.2290}, electrical resistivity, magnetic susceptbility, M\"{o}ssbaurer spectroscopy and single crystal neutron scattering experiments~\cite{JMMM.132.46}.
In spite of this, the small magnetic moments detected by M\"{o}ssbaurer spectroscopy and single crystal neutron scattering experiments still manifest partially itinerant 5$f$ states. That is to say, the partially itinerant or localized 5$f$ electrons of NpSn$_3$ is still perplexing. In analogy with USn$_3$, the hybridzation of 5$f$ states with conduction bands greatly affects the localization degree of freedom and heavy-fermion behavior. On the other hand, limited theoretical studies could not elucidate the itinerant antiferromagnetic ordering using crystal field splitting model~\cite{PhysRevLett.65.2290} and fail to accurately delineate the strongly correlated 5$f$ electrons~\cite{JMMM320.503,KHAN201877}.
It is reported that PuSn$_3$ shows temperature-independent paramagnetism~\cite{Handbookferromag,PhysRevB.39.13115} and pseudogap around the Fermi level in low-temperature electrical resistivity~\cite{Brodsky1978}, still lacking electronic structure characterization experiments such as photoemission spectroscopy, angle-resolved photoemission spectroscopy, and x-ray adsorption spectroscopy. Meanwhile, theoretical investigations have made great efforts to systematically obtain the crystal structure, bulk modulus, electronic structure, quantum oscillation~\cite{PhysRevB.39.13115,BAIZAEE2005247,BAIZAEE2007287,PhysRevB.88.125106}, leaving the underlying physics of paramagnetic ground state being untouched.  

Actually, plentiful theoretical efforts have been devoted to study uranium pnictides and chalcogenides~\cite{PhysRevB.12.4102,PhysRevB.17.3104}. The band structure, density of states, fermi surface and electron effective mass of USn$_3$ have been previously investigated by employing first-principles calculations~\cite{PhysRevB.59.14473,Strange1986}. Subsequently, the strong correlation among 5$f$ electrons and large spin-orbit coupling complicate the calculation of $An$Sn$_3$ ($An$=U, Np, Pu) compounds, blurring a comprehensive physical picture of electronic structure and ground state properties.
Evidently, the missing narrow flat 5$f$ bands within the traditional density functional theory is attributed to the underestimation of strongly correlated 5$f$ electrons. Furthermore, the spin-orbit coupling of actinides should be carefully treated. 
Especially, the intrinsic mechanism for the paramagnetic ground state of USn$_3$ and NpSn$_3$ remains elusive even though the spin-fluctuation theory has been proposed for USn$_3$. Additionally, the bewildering heavy-fermion behavior of $An$Sn$_3$ ($An$=U, Np, Pu) is rarely scrutinized. Now that the hybridization strength between 5$f$ states and conduction bands greatly influences the low temperature magnetic ordering and itinerant-localized dual nature of 5$f$ electrons, it is illuminating to investigate the 5$f$ series dependence of 5$f$ electronic structure and 5$f$ electron hydridization with ligand states to uncover the profound bonding behavior and strongly correlated 5$f$ states, which shall gain deep insight on the diverse ground state ordering as well as alluring heavy-fermion behavior.

\begin{table*}[ht]
\caption{Crystal structure parameters, electronic specific heat coefficient and ground state ordering of $An$Sn$_3$ ($An$=U, Np, Pu). \label{tab:param}}
\begin{tabular}{ccccc}
\hline\hline
cases & 
$d_{An-An}$ (\r{A}) &
$d_{An-Sn}$ (\r{A}) &
$\gamma$ (mJ/(mol$\times$K$^2$)) &
ground state \\
\hline
USn$_3$  & 4.603 & 3.255 & 171\footnotemark[1] & paramagnetism\footnotemark[4] \\

NpSn$_3$ & 4.627 & 3.272 & 242\footnotemark[2] & itinerant-electron antiferromagnetism\footnotemark[2] (T$_N$ = 9.5 K) \\

PuSn$_3$ & 4.630 & 3.274 & 96\footnotemark[3] & paramagnetism\footnotemark[5] \\
\hline\hline
\end{tabular}
\footnotetext[1]{See Ref.~[\onlinecite{PhysRevB.33.8035}].}
\footnotetext[2]{See Ref.~[\onlinecite{PhysRevLett.37.1511}].}
\footnotetext[3]{See Ref.~[\onlinecite{LuPuSn3}].}
\footnotetext[4]{See Ref.~[\onlinecite{PhysRevB.41.9294}].}
\footnotetext[5]{See Ref.~[\onlinecite{PhysRevB.39.13115}].}
\end{table*}


\section{Methods\label{sec:method}}
In order to capture the essence in the strongly correlated 5$f$ electrons for uranium, neptonium and plutonium, 5$f$ correlated electronic states dependence on actinide elements are investigated by means of a combination of the density functional theory and the embedded dynamical mean-field theory (DFT + DMFT). The DFT + DMFT method integrates the DFT realistic band structure calculation and a non-perturbative way to tackle the many-body local interaction effects in DMFT~\cite{RevModPhys.68.13,RevModPhys.78.865}. Here the strong electronic correlation and large spin-orbit coupling are treated on an equal footing in the calculation. Then the DFT + DMFT approach is implemented into DFT and DMFT parts, which are solved separately by using the \texttt{WIEN2K} code~\cite{wien2k} and the \texttt{EDMFTF} software package~\cite{PhysRevB.81.195107}.  

In the DFT calculation, the experimental crystal structures of $An$Sn$_3$ ($An$=U, Np, Pu) were utilized. Since the calculated temperature 116 K is above the antiferromagnetic transition temperature of NpSn$_3$, the system was assumed to be nonmagnetic. The generalized gradient approximation was tested and used to formulate the exchange-correlation functional~\cite{PhysRevLett.77.3865}. Since the actinides belong to heavy element, the considerable spin-orbit coupling was treated in a second-order variational manner. The $k$-points' mesh was $15 \times 15 \times 15$ and $R_{\text{MT}}K_{\text{MAX}} = 8.0$.  

In the DMFT part, 5$f$ electrons of $An$ atom were treated as correlated. The four-fermions' interaction matrix was parameterized using the Coulomb interaction $U = 5.0$~eV and the Hund's exchange $J_H=0.6$~eV via the Slater integrals~\cite{PhysRevB.59.9903} for $An$Sn$_3$ ($An$=U, Np, Pu). It should be noted that the band structure and density of states for $U = $ 4.0~eV, 5.0~eV, 6.0~eV reveal subtle nuance for $An$Sn$_3$ ($An$=U, Np, Pu). The fully localized limit scheme was used to compute the double-counting term for impurity self-energy function~\cite{jpcm:1997}. The vertex-corrected one-crossing approximation (OCA) impurity solver~\cite{PhysRevB.64.115111} was employed to solve the resulting multi-orbital Anderson impurity models. In order to improve the computation burden, some good quantum numbers such as $N$ and $J$ were chosen to divide the whole Hilbert space~\cite{PhysRevB.75.155113} into subblocks. Moreover, the truncation approximation for the atomic eigenstates was adopted to gain further acceleration. Explicitly, only those atomic eigenstates whose occupancy $N$ satisfy $N$ $\in$ $[N_{low}, N_{high}]$ will be taken into account in the local trace evaluation. Finally, the convergence criteria for energy and charge were $10^{-5}$ Ry and $10^{-5}$ e, respectively. It is worth mentioning that the direct output of OCA impurity solver is real axis self-energy $\Sigma (\omega)$ which was used to calculate the momentum-resolved spectral functions $A(\mathbf{k},\omega)$, density of states $A(\omega)$ and other physical observables.

\section{Results\label{sec:results}}

\begin{figure*}[th]
\centering
\includegraphics[width=\textwidth]{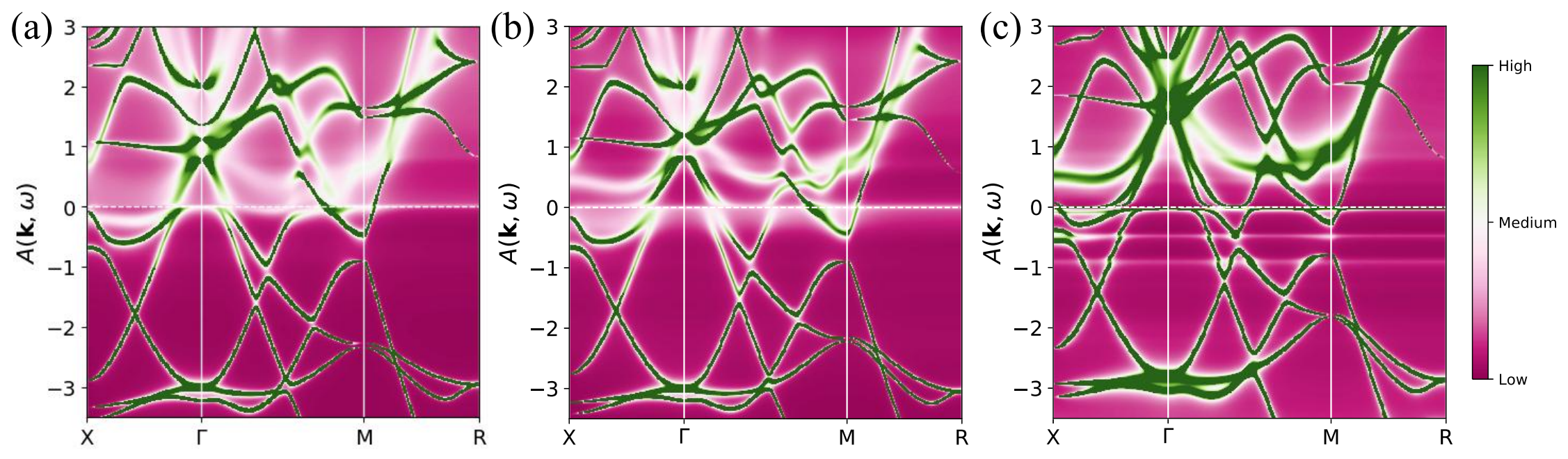}
\caption{(Color online). Momentum-resolved spectral functions $A(\mathbf{k},\omega)$ of $An$Sn$_3$ ($An$=U, Np, Pu) at 116 K under ambient pressure calculated by the DFT + DMFT method. The horizontal lines denote the Fermi level. (a) USn$_3$. (b) NpSn$_3$. (c) PuSn$_3$. 
\label{fig:akw}}
\end{figure*}

\begin{figure*}[th]
\centering
\includegraphics[width=0.9\textwidth]{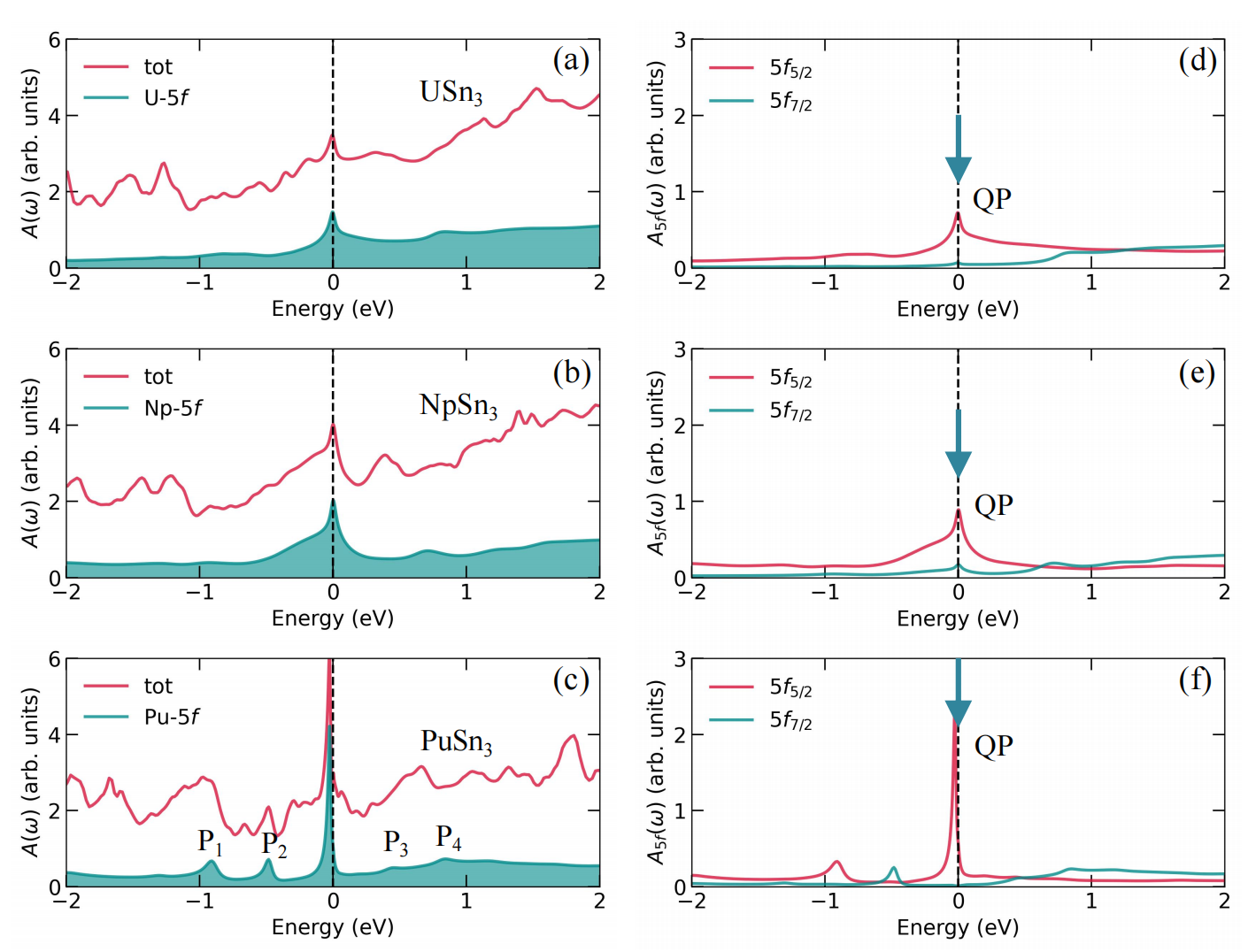}
\caption{(Color online). Electronic density of states of $An$Sn$_3$ ($An$=U, Np, Pu) at 116 K obtained by the DFT + DMFT method. Total density of states (thick solid lines) and partial 5$f$ density of states (color-filled regions) of (a) USn$_3$, (b) NpSn$_3$, (c) PuSn$_3$. These peaks in figure (c) resulting from the quasiparticle multiplets are denoted with ``P1'', ``P2'', ``P3'', and ``P4''. The $j$-resolved 5$f$ partial density of states with $5f_{5/2}$ and $5f_{7/2}$ components represented by red and green lines, respectively. (d) USn$_3$, (e) NpSn$_3$, (f) PuSn$_3$. 
\label{fig:dos}}
\end{figure*}

\begin{figure*}[th]
\centering
\includegraphics[width=\textwidth]{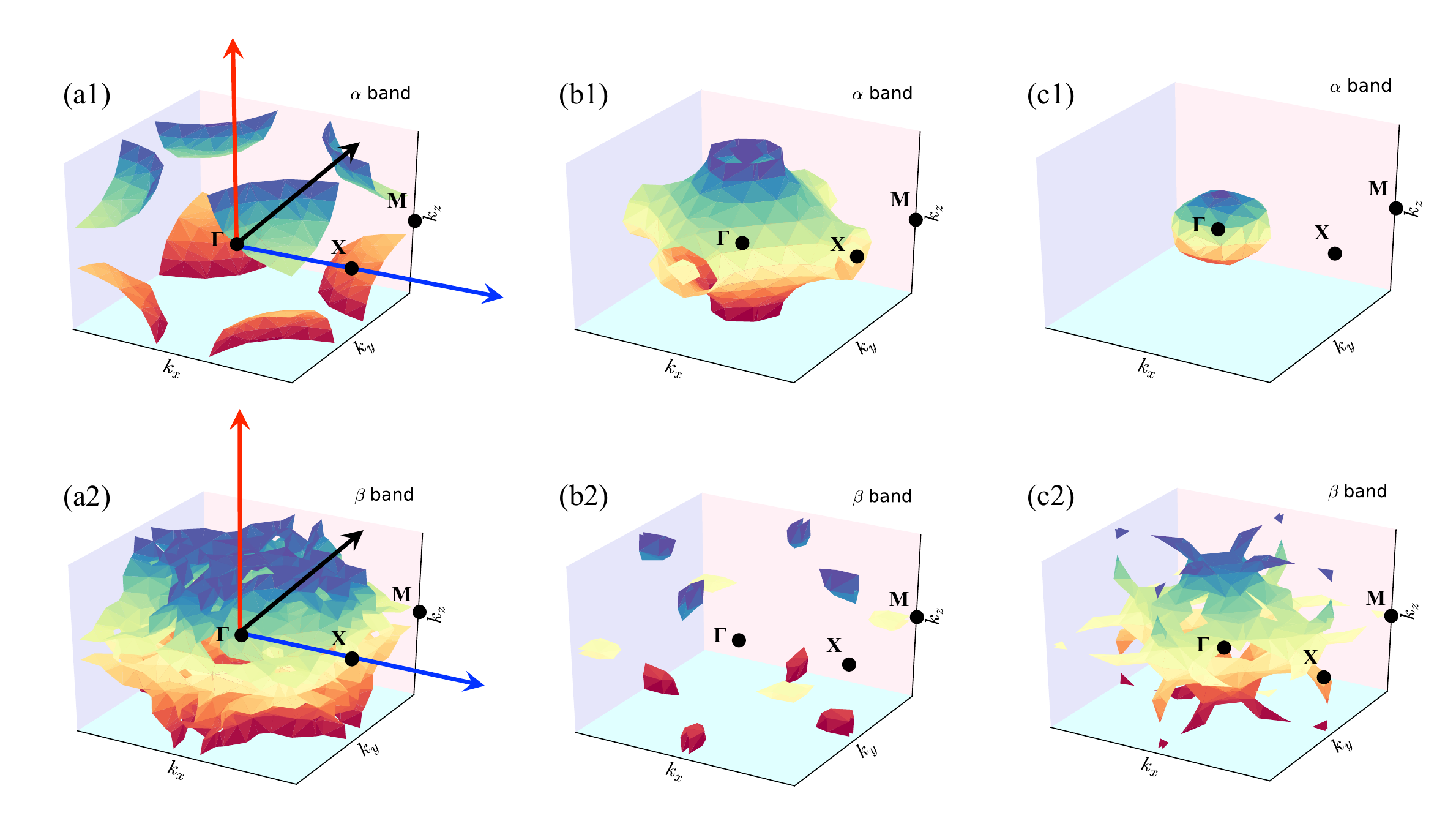}
\caption{(Color online). Three-dimensional Fermi surface of $An$Sn$_3$ ($An$=U, Np, Pu) at 116 K calculated by the DFT + DMFT method. (a1)-(a2) USn$_3$. (b1)-(b2) NpSn$_3$. (c1)-(c2) PuSn$_3$. 
\label{fig:3dFS}}
\end{figure*}

\begin{figure*}[th]
\centering
\includegraphics[width=\textwidth]{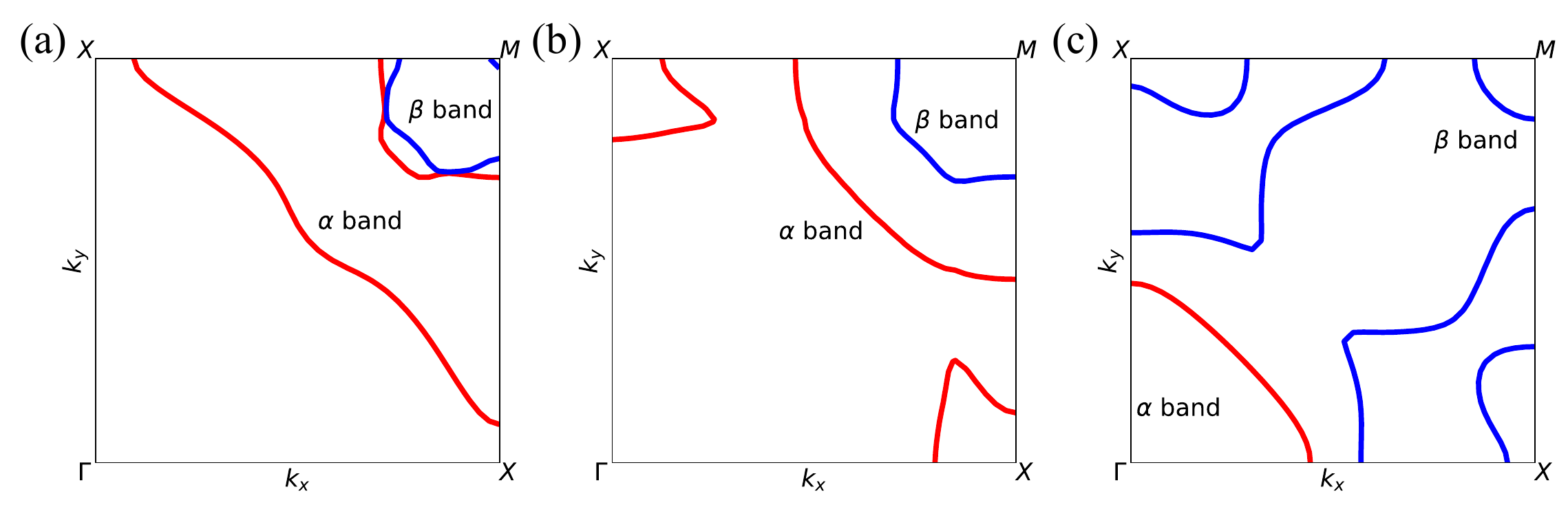}
\caption{(Color online). Two-dimensional Fermi surface on the $k_x-k_y$ plane (with $k_z = \pi/2$) of $An$Sn$_3$ ($An$=U, Np, Pu) at 116 K calculated by the DFT + DMFT method. (a) USn$_3$. (b) NpSn$_3$. (c) PuSn$_3$. They are visualized with different colors.
\label{fig:2dFS}}
\end{figure*}

\subsection{Electronic band structures}
Here we concentrate on the momentum-resolved spectral functions $A(\mathbf{k},\omega)$ of $An$Sn$_3$ ($An$=U, Np, Pu) to reveal the inherent nature of 5$f$ electrons. To verify the reliability of our calculations, we examine the electronic band structure of USn$_3$ along high-symmetry line $X - \Gamma - M - R $ in the irreducible Brillouin zone [see Fig.~\ref{fig:akw}(a)]. Compared with the available theoretical results by A. L. Cornelius {\it et al.} employing a localized spin density approximation~\cite{PhysRevB.59.14473}, the essential feature of band structures [see Fig.~\ref{fig:akw}(a)] generally coincide with each other. It is noticed that the hole-like orbit around the Fermi level at $\Gamma$ point and similar hole-like orbits at $X$ and $M$ points are almost alike. However, the major difference is that flat narrow 5$f$ electronic bands intersect the conduction bands in the vicinity of the Fermi level and open evident hybridization gaps. It is proposed that the absence of quasiparticle 5$f$ bands in literature~\cite{PhysRevB.59.14473,Strange1986} is probably concerned with the underestimation of strongly correlated 5$f$ electrons without sophisticated treating the many-body effects. 

Then we scrutinize the salient traits encoded in the band structure of $An$Sn$_3$ ($An$=U, Np, Pu). It is observed that the overall profile of the momentum-resolved spectral functions for USn$_3$ and NpSn$_3$ [see Fig.~\ref{fig:akw}(a) and~\ref{fig:akw}(b)] at 116 K share abundant similarities with analogous conduction orbits and remarkable dispersionless 5$f$ electronic bands around the Fermi level. Even so, a slight discrepancy gives a higher 5$f$ band intensity of NpSn$_3$ than that of USn$_3$, concretising the itinerant instinct of 5$f$ electrons. In comparison with the band structure of PuSn$_3$ [see Fig.~\ref{fig:akw}(c)], it is found out that a mild difference in the conduction bands above the Fermi level and almost the same conduction bands below the Fermi level contributed by 5$p$ states of Sn atom. In particular, the apparent narrow 5$f$ bands emerge at -0.9 eV, -0.47 eV and the Fermi level, which hybridize with ligand states to form obvious hybridization gaps. Correspondingly, prominent spectral weight of 5$f$ electrons, dramatic quasiparticle bands and striking hybridization entirely hint the itinerant and coherent 5$f$ states of PuSn$_3$. Then it is speculated that the distinct character of 5$f$ electrons roots from various actinide series and various hybridization strength bewteen 5$f$-ligand states.

\subsection{Density of states\label{sec:DOS}}
To unveil the 5$f$ electronic structure of $An$Sn$_3$ ($An$=U, Np, Pu), it is enlightening to explore the density of states and quasiparticle peaks depicted in Figs.~\ref{fig:dos}(a)-(g). Firstly, the entire outline of total density of states for USn$_3$ and NpSn$_3$ bears striking resemblances including a sharp narrow quasiparticle peak at the Fermi level and a broad ``hump" in the energy region of 0.5 eV $\sim$ 2 eV. The prominent quasiparticle peaks in the Fermi level delineate the itinerancy and coherence of 5$f$ electrons.
Owing to the spin-orbit coupling, the fourteen-fold degenerated 5$f$ orbitals are split into six-fold degenerated 5$f_{5/2}$ and eight-fold degenerated 5$f_{7/2}$ subbands [see Figs.~\ref{fig:dos}(d)-(f)]. The archetypical quasiparticle peak is mostly contributed by 5$f_{5/2}$ orbital of USn$_3$ and NpSn$_3$. Meanwhile, a slight increase in the spectral weight of 5$f$ orbital of NpSn$_3$ with respect to USn$_3$ may attributed a visible spectral weight of 5$f_{7/2}$ orbital in the Fermi level of NpSn$_3$. As is listed in table~\ref{tab:sig}, the quasi-particle weight $Z$ of 5$f_{7/2}$ orbital in NpSn$_3$ is maximal which demonstrates the substantial valence state fluctuations and mixed-valence behavior.

Interestingly, the most distinctive feature of the density of states of PuSn$_3$ [see Fig.~\ref{fig:dos}(c)] is the typical atomic multiplets with one keen-edged quasiparticle peak in the Fermi level and four satellite peaks above and below the Fermi level. In combination with the partial 5$f$ density of states [see Fig.~\ref{fig:dos}(f)], the quasiparticle peak is contributed by 5$f_{5/2}$ orbital. Meanwhile, the two satellite peaks ``P1'' and ``P2'' at --0.9 eV and --0.47 eV with energy gap about 0.43 eV are attributed to 5$f_{5/2}$ and 5$f_{7/2}$ orbitals, respectively. Above the Fermi level, the reflected peaks ``P3'' and ``P4'' distribute in 0.47 eV and 0.9 eV relative to the central quasiparticle. Thus the five representative peaks are called ``quasiparticle multiplets'' which commonly exist in plutonium-based compounds such as Pu$_3$Ga, PuIn$_3$ and PuB$_6$~\cite{PhysRevB.103.205134,PuBx34.215601}. To trace the origin of these quasiparticle multiplets, it is deduced that ``P3'' and ``P4'' peaks are formed by a mix of 5$f_{5/2}$ and 5$f_{7/2}$ orbitals. It is mentioned that the quasiparticle multiplets are induced by abundant competing atomic eigenstates of 5$f$ electrons synergy with 5$f$ valence state fluctuations discussed in subsection~\ref{sec:valence}.
In other words, ample competitive atomic eigenstates act as prerequisites for the onset of quasiparticle multiplets. For instance, nominal occupancy of 5$f$ electrons of Pu atom is 5.0, which is larger than those of uranium and neptunium 2.0 and 3.0, respectively. Then the number of occupied 5$f$ states of Pu atom is larger than those of uranium and neptunium. In contrast, the enhancing localization of 5$f$ electrons for curium and californium with relatively weak valence state fluctuations fail to produce quasiparticle multiplets even though their nominal occupancy of 5$f$ electrons is larger than 5.0. That is why only PuSn$_3$ develops observable quasiparticle multiplets even if valence state fluctuations are remarkable in $An$Sn$_3$ ($An$=U, Np, Pu).

\subsection{Fermi surface topology}
Fermi surface topology is a pivotal physical quantity to character the detail electronic structure of $An$Sn$_3$ ($An$=U, Np, Pu). Figure~\ref{fig:3dFS} visualizes the three-dimensional Fermi surface topology at 116 K. 
There exist two doubly degenerated bands crossing the Fermi level (No. of bands: 16 and 17, 18 and 19) for USn$_3$ and (No. of bands: 18 and 19, 20 and 21) for both NpSn$_3$ and PuSn$_3$, which are labeled by $\alpha$ and $\beta$, respectively. For USn$_3$ [see Fig.~\ref{fig:3dFS}(a1) and (a2)], $\alpha$ band occupies eight corners of the first Brillouin zone and $\beta$ band resembles an anisotropic huge form. Especially, $\alpha$ band develops a distorted hexahedral structure for NpSn$_3$ [see Fig.~\ref{fig:3dFS}(b1) and (b2)], while $\beta$ band displays twelve small scattered fragments in the edge centers of the first Brillouin zone.
Note that for PuSn$_3$ [see Fig.~\ref{fig:3dFS}(c1) and (c2)] the Fermi surface topology of $\alpha$ band takes an ellipsoid shape, which agrees with those in previous DFT calculations~\cite{PhysRevB.88.125106}. Nevertheless, the Fermi surface topology of $\beta$ band forms an anisotropic shape, which acts distinctively from the previous results~\cite{PhysRevB.88.125106}. The discrepancy of $\beta$ band might originate from the temperature effect and the strongly correlated 5$f$ electronic states on Fermi surface. The DFT calculations~\cite{PhysRevB.88.125106} are performed at zero temperature which also underestimates the 5$f$ electronic correlation. Besides, it is pointed out that the Fermi surface could be detected by following de Haas–van Alphen quantum oscillation, then the experimental results may clarify possible physical causes behind the paramagnetic ground state of USn$_3$ and PuSn$_3$ provided that no Fermi surface nesting is observed.
 
Since the interior structure of three-dimensional Fermi surface is hard to discern, the two-dimensional Fermi surface of $\alpha$ and $\beta$ bands are plotted in Fig.~\ref{fig:2dFS}. Corresponding to the three-dimensional Fermi surface of USn$_3$ [see Fig.~\ref{fig:2dFS}(a1) and (a2)], both $\alpha$ and $\beta$ bands intersect the $\Gamma$ - $M$ line and $M$ - $X$ line, while no bands cross the $\Gamma$ - $X$ line.  
Meanwhile, only $\alpha$ band cuts through the $\Gamma$ - $X$ line for NpSn$_3$ [see Fig.~\ref{fig:2dFS}(b1) and (b2)] with $\alpha$ and $\beta$ bands passing through the $\Gamma$ - $M$ line and $M$ - $X$ line.
Finally, both $\alpha$ and $\beta$ bands cross the $\Gamma$ - $X$ line and $\Gamma$ - $M$ line for PuSn$_3$ [see Fig.~\ref{fig:2dFS}(c1) and (c2)], along with $\beta$ band intersecting $M$ - $X$ line, which is in accordance with the momentum-resolved spectral functions [see Fig.~\ref{fig:akw}].

\subsection{Valence state fluctuations\label{sec:valence}}

\begin{table}[th]
\caption{Probabilities of 5$f^n$ electronic configuration at 116 K for USn$_3$ ($n \in$ [0, 4]), NpSn$_3$ ($n \in$ [1, 5]) and PuSn$_3$ ($n \in$ [3, 7]), 5$f$ occupancy $n_{5f}$, x-ray absorption branching ratio $\mathcal{B}$. \label{tab:prob}}
\begin{ruledtabular}
\begin{tabular}{cccccccc}
USn$_3$ & $5f^{0}$ & $5f^{1}$ & $5f^{2}$ & $5f^{3}$ & $5f^{4}$ & $n_{5f}$ & $\mathcal{B}$ \\
  & 9.048$\times 10^{-5}$ & 0.026 & 0.671 & 0.297 & 0.012 & 2.283 & 0.6855 \\
\hline
NpSn$_3$ & $5f^{1}$ & $5f^{2}$ & $5f^{3}$ & $5f^{4}$ & $5f^{5}$ & $n_{5f}$ & $\mathcal{B}$\\
 & 9.234$\times 10^{-5}$ & 0.018 & 0.472 & 0.496 & 0.014 & 3.506 & 0.7291 \\
\hline
PuSn$_3$ & $5f^{3}$ & $5f^{4}$ & $5f^{5}$ & $5f^{6}$ & $5f^{7}$ & $n_{5f}$ & $\mathcal{B}$\\
  & 9.976$\times 10^{-4}$ & 0.075 & 0.772 & 0.150 & 1.343$\times 10^{-3}$ & 5.076 & 0.7806 \\
\end{tabular}
\end{ruledtabular}
\end{table}

Now that alluring quasiparticle multiplets are closely interrelated with valence state fluctuations, the 5$f$ electron atomic eigenstates obtained from the output of DMFT many-body states contain the valence state fluctuations. $p_\Gamma$ is used to quantify the probability of 5$f$ electrons which remains in each atomic eigenstate $\Gamma$. Then the average 5$f$ valence electron is defined as $n_{5f} = \sum_\Gamma p_\Gamma n_\Gamma$, where $n_\Gamma$ denotes the number of electrons in each atomic eigenstate $\Gamma$. Finally, the probability of 5$f^n$ electronic configuration can be written as $P_n = w(5f^{n}) = \sum_\Gamma p_\Gamma \delta (n-n_\Gamma)$. 

The calculated probabilities of 5$f^n$ electronic configuration for $An$Sn$_3$ ($An$=U, Np, Pu) are listed in table~\ref{tab:prob}. Evidently, for USn$_3$, the probability of 5$f^2$ electronic configuration accounts for 67.1\%, followed by the probability of 5$f^3$ about 29.7\%, giving the probabilities of 5$f^1$ and 5$f^4$ electronic configurations about 9.1\% and 1.2\%, respectively. The conspicuous valence state fluctuations give rise to the 5$f$ electron occupancy $n_{5f}$ = 2.283, which deviates from its nominal value 2.0. Accordingly, 5$f$ electrons tend to spend more time on 5$f^2$ and 5$f^3$ electronic configurations, promoting mixed-valence behavior. Referring to NpSn$_3$, the competitive probabilities of 5$f^3$ and 5$f^4$ electronic configurations approximate 47.2\% and 49.6\%, respectively, which render substantial valence state fluctuations. Together with a small portion of the probabilities of 5$f^2$ and 5$f^5$ electronic configurations around 1.8\% and 1.4\%, the 5$f$ electron occupancy $n_{5f}$ = 3.506 diverges from its nominal value 3.0 to a large extent. Therefore NpSn$_3$ exhibits noteworthy valence state fluctuations and related mixed-valence states. Eventually, we focus on the probability of electronic configuration of PuSn$_3$. The leading probability of 5$f^5$ electronic configuration reaches as high as 77.2\%. Meanwhile, the probability of 5$f^6$ and 5$f^4$ attain 15.0\% and 7.5\%, respectively, neglecting other electronic configurations. In this context, a moderate valence state fluctuations regulate the 5$f$ electron occupancy $n_{5f}$ = 5.076, which approaches its nominal value 5.0. 
In order to quantitatively describe the strength of valence state fluctuations, we define the quantity ${\cal V}=\sum_n P_n(P_n-1)$, where $n$ denotes the occupancy number of 5$f$ electrons. Two extreme cases are listed for illustration. The case for $P_n = 1$, ${\cal V} = 0$ signifies the absence of valence state fluctuations with only one electronic configuration. In this sense, electrons are extremely localized where they prefer to stay in one electronic configuration. On the contrary, dispersive distribution of electronic configurations generates arresting valence state fluctuations with ${\cal V} = 1$. Actually, such ideal systems hardly exist in real materials and ${\cal V}$ evaluates 0.689, 0.623, 0.449 for USn$_3$, NpSn$_3$, PuSn$_3$, respectively. Consequently, the strength of valence state fluctuations obeys the following sequence, ${\cal V}$(USn$_3)$ $\textgreater$ ${\cal V}$(NpSn$_3)$ $\textgreater$ ${\cal V}$(PuSn$_3)$. The order conforms with the qualitative picture that increscent atomic number with growing localization of 5$f$ electrons suppresses valence state fluctuations. It is expected that ongoing research on a wide range of $f$ electronic states for lanthanide and actinide compounds shall shed light on the comprehensive results about precise description of valence state fluctuations.

\subsection{Self-energy functions\label{sec:selfenergy}}

\begin{table}[t]
\caption{The effective electron mass $m^\star$ and quasi-particle weight $Z$ of $5f_{5/2}$ and $5f_{7/2}$ states for $An$Sn$_3$ ($An$=U, Np, Pu) at 116 K. \label{tab:sig}}
\begin{ruledtabular}
\begin{tabular}{ccccc}
& \multicolumn{2}{c}{$5f_{5/2}$} & \multicolumn{2}{c}{$5f_{7/2}$} \\
compounds & $m^{\star}/m_e$ & $Z$ & $m^{\star}/m_e$ & $Z$ \\
\hline
USn$_3$  & 18.717  & 0.053 & 17.233 & 0.058 \\
NpSn$_3$ & 18.673 & 0.054 & 1.890  & 0.529 \\
PuSn$_3$ & 29.111 & 0.034 & 13.873 & 0.072 \\
\end{tabular}
\end{ruledtabular}
\end{table}

Inspiringly, the electronic correlations are encoded in the electron self-energy functions~\cite{RevModPhys.68.13,RevModPhys.78.865}. The renormalization factor $Z$ which embodies the electronic correlation strength can be deduced from the real part of self-energy functions via the following equation~\cite{RevModPhys.68.13}:
\begin{equation}
Z^{-1} = \frac{m^\star}{m_e} = 1 - \frac{\partial \text{Re} \Sigma(\omega)}{\partial \omega} \Big|_{\omega = 0}. \label{eqsigma}
\end{equation}

Here renormalization factor $Z$ quantitatively depicts the quasiparticle weight and band compression degree of correlated electronic bands which is inversely proportional to the electron effective mass $m^\star$. Table~\ref{tab:sig} lists the computed effective electron mass $m^\star$ and renormalization factor $Z$ of $5f_{5/2}$ and $5f_{7/2}$ states for $An$Sn$_3$ ($An$=U, Np, Pu) at 116 K. 
For USn$_3$, sizable electron effective mass for both $5f_{5/2}$ and $5f_{7/2}$ states implicates the strongly correlated 5$f$ electronic states, with suppressed quasiparticle weight $Z$. Similar to USn$_3$, PuSn$_3$ reveals strong correlation strength among 5$f$ states where the electron effective mass of $5f_{5/2}$ state is much larger than $5f_{7/2}$ state. As mentioned in subsection~\ref{sec:DOS}, the finite distribution of spectral weight of $5f_{7/2}$ state of NpSn$_3$ arises from the smooth variation across the Fermi level, which indicates the small electron effective mass of $5f_{7/2}$ state with significant quasi-particle weight $Z$. The comparatively weak electron correlation of $5f_{7/2}$ state compared to $5f_{5/2}$ state, evincing the orbital dependent character. It is speculated that the orbital selective correlation of $5f_{5/2}$ and $5f_{7/2}$ states for NpSn$_3$ holds underlying connection with itinerant antiferromagnetic ordering.

\section{Discussion\label{sec:dis}}
In this section, we discuss the hybridization between 5$f$ states and conduction bands to reveal the paramagnetic ground state of USn$_3$ and PuSn$_3$, as well as the antiferromagnetic ordering of NpSn$_3$. Moreover, we would like to interpret the angular momentum coupling scheme through x-ray absorption spectroscopy.

\textbf{5$f$ electron dependence of electronic structure.}
As atomic number increases from uranium to plutonium, the itinerant-localized 5$f$ states interrelates with 5$f$ electron occupancy. As mentioned above, the spectral weight of 5$f$ states near the Fermi level demonstrates that 5$f$ electrons are inclined to hybridize with Sn-5$p$ bands to form partially itinerant 5$f$ states. The bonding behavior is also evidenced in the significant valence state fluctuations and apparent mixed-valence behavior for $An$Sn$_3$ ($An$=U, Np, Pu). It is speculated that larger atomic actinide distance may result in weaker $f-c$ hybridization so as to suppress itinerant degree of freedom and restrain valence state fluctuations. Unexpectedly, quasiparticle multiplets in PuSn$_3$ is prominent which is intimately associated with valence state fluctuations and multiple competing atomic eigenstates. The distinctive feature of PuSn$_3$ restates that Pu sits on the edge of itinerant-localized 5$f$ states. In this sense, USn$_3$ and NpSn$_3$ tend to develop paramagnetic ground state because of substantial $f-c$ hybridization. Even though NpSn$_3$ exhibits antiferromagnetic ordering, the magnetic moment of Np atom is small. This is mainly attributed to the two competitive electronic configurations 5$f^3$ and 5$f^4$ which renders noteworthy valence state fluctuations and mixed-valence nature.

\textbf{Angular momentum coupling scheme.}
For the actinides with intricate spin-orbit coupling, 5$f$ electrons occupy the $5f_{5/2}$ and $5f_{7/2}$ subbands. Since the evolution trend of 5$f$ electron in the $5f_{5/2}$ and $5f_{7/2}$ levels with respect to incremental atomic radius is a vital problem, the angular momentum coupling scheme serves as the primary determinant factor. 
For the multielectronic systems, there exist three ways of angular momentum coupling in light of the relative strength of spin-orbit coupling and electronic interaction, including Russell-Saunders (LS) coupling, $jj$ coupling, and intermediate coupling (IC)~\cite{RevModPhys.81.235}. For the ground state of heavy actinides, intermediate coupling scheme is usually the favorite.
A natural question rises with the 5$f$ orbital occupancy and angular momentum coupling scheme for $An$Sn$_3$ ($An$=U, Np, Pu).
 
Since x-ray absorption spectroscopy is a powerful technique to detect the electronic transitions between core 4$d$ and valence 5$f$ states, it is employed to observe the occupancy of 5$f$ electrons of actinides. The strong spin-orbit coupling for the 4$d$ states leads to two absorption lines, representing the $4d_{5/2} \rightarrow 5f$ and $4d_{3/2} \rightarrow 5f$ transitions, respectively. The x-ray absorption branching ratio $\mathcal{B}$ is defined as the relative strength of the $4d_{5/2}$ absorption line~\cite{PhysRevA.38.1943}. It calibrates the spin-orbit coupling interaction strength in 5$f$ shell. Under the approximation that the electrostatic interaction between core and valence electrons is ignored, the expression for $\mathcal{B}$ is written as~\cite{shim:2007}:
\begin{equation}
\label{eq:ratio}
\mathcal{B} = \frac{3}{5} - \frac{4}{15} \frac{1}{14 - n_{5/2} - n_{7/2}} \left ( \frac{3}{2} n_{7/2} - 2 n_{5/2} \right ),
\end{equation}
where $n_{7/2}$ and $n_{5/2}$ are the 5$f$ occupation numbers for the $5f_{7/2}$ and $5f_{5/2}$ states, respectively. The computed results are listed in table~\ref{tab:prob}, obeying $\mathcal{B}$(PuSn$_3$) $\textgreater$ $\mathcal{B}$(NpSn$_{3}$) $\textgreater$ $\mathcal{B}$(USn$_3$). The sequence evinces that the angular momentum coupling scheme of three compounds belongs to the intermediate coupling scheme, which is analogous to that in heavy actinides Pu~\cite{PhysRevB.101.125123} and Cm~\cite{PhysRevB.101.195123}.


\section{conclusion\label{sec:summary}}

In summary, we employ the density functional theory in combination with embedded dynamical mean-field theory to study extensively the 5$f$ strongly correlated electronic states of isostructural compounds $An$Sn$_3$ ($An$=U, Np, Pu). It is found that apparent quasiparticle weight of 5$f$ states around the Fermi level is accompanied by dispersionless  5$f$ electronic band. Even though quasiparticle multiplets only develop in PuSn$_3$, striking valence state fluctuations imply the itinerancy of 5$f$ electrons and the tendency to hybridize with ligand states to facilitate paramagnetic ground state.
Usually, localized 5$f$ states interrelate with strongly correlated 5$f$ electrons with renormalized bands and large electron effective mass. Then 5$f$ electrons are strongly correlated in three compounds with orbital selective correlation of $5f_{7/2}$ and $5f_{5/2}$ states for NpSn$_3$. Additionally, the angular momentum coupling scheme is found to be intermediate which could be utilized to detect the 5$f$ electron occupancy and related physical properties. Briefly, the evolution pattern of strongly correlated 5$f$ electron and itinerant-localized 5$f$ states dependence on actinide series gain valuable implications into the actinide compounds and pave the way for future research.

\begin{acknowledgments}
This work was supported by National Natural Science Foundation of China 2022YFA140220, CAEP Project (No.~TCGH0710) and the National Natural Science Foundation of China (No.~11874329, No.~11934020). 
\end{acknowledgments}


\bibliography{us}

\end{document}